\documentclass[twocolumn,superscriptaddress,nofootinbib,aps,prl,floatfix,preprintnumbers,amsmath,amssymb,groupedaddress]{revtex4}
\usepackage{dsfont}
\usepackage{epsfig}
\usepackage{slashed}
\usepackage{bbold}
\usepackage{psfrag}
\usepackage{color}
\PassOptionsToPackage{caption=false}{subfig}
\usepackage{subfig}
\usepackage{multirow}
\usepackage{booktabs}
\usepackage{pstricks}
\usepackage{feynmp}
\usepackage{color}
\usepackage{hyperref}
\usepackage{epsfig,amsmath,amssymb,verbatim,mathrsfs,array,layout,textcomp,amssymb,latexsym}

\newenvironment{Eqnarray}{\arraycolsep 0.14em\begin{eqnarray}}{\end{eqnarray}}
\def\beqa{\begin{Eqnarray}}
\def\eeqa{\end{Eqnarray}}
\def\beq{\begin{equation}}
\def\eeq{\end{equation}}
\newcommand{\no}\nonumber

\newcommand{\rf}{r_f}

\definecolor{red1}{cmyk}{0,1,1,0.3}

\begin{document}
\title{Constraining the Higgs-Dilaton with  LHC and Dark Matter Searches}
\author{Aielet Efrati}
\affiliation{Department of Particle Physics and Astrophysics, Weizmann Institute of Science, Rehovot 7610001, Israel}
\author{Eric Kuflik}
\affiliation{Raymond and Beverly Sackler School of Physics and Astronomy, Tel-Aviv University, Tel-Aviv 69978, Israel}
\affiliation{Department of Physics, LEPP, Cornell University, Ithaca, NY 14853}
\author{Shmuel Nussinov}
\affiliation{Raymond and Beverly Sackler School of Physics and Astronomy, Tel-Aviv University, Tel-Aviv 69978, Israel}
\author{Yotam Soreq}
\affiliation{Department of Particle Physics and Astrophysics, Weizmann Institute of Science, Rehovot 7610001, Israel}
\author{Tomer Volansky}
\affiliation{Raymond and Beverly Sackler School of Physics and Astronomy, Tel-Aviv University, Tel-Aviv 69978, Israel}

\begin{abstract}
We study a scenario in which the dilaton, a pseudo-Goldstone boson of the spontaneous breaking of conformal symmetry, provides a portal between dark matter and the visible sector. We consider the low-energy description of the theory in which the dilaton mixes with the Standard Model Higgs boson, thereby predicting a second scalar at or above the weak scale. We derive the collider and dark matter constraints on the corresponding parameter space and find that existing experimental data point towards the decoupling limit in which the CFT scale is well above the electroweak scale.  Moreover, the thermal production of dark matter implies its mass is likely above the TeV scale. Upcoming direct detection experiments may allow for the discovery of the dilaton-mediated thermal dark matter while future collider studies will also be sensitive to the available parameter space.
\end{abstract}
\maketitle
%

\section{1. Introduction}
\label{sec:intro}

The recently discovered particle at the LHC, with mass in the vicinity of 125~GeV~\cite{Aad:2012tfa,Chatrchyan:2012ufa}, has properties which closely resemble those of the Standard Model (SM) Higgs boson~\cite{ATLAS:2014, CMS:2014}. Nevertheless, it is still possible that this particle is an impostor, not directly or entirely related to the breaking of the electroweak symmetry. A motivated example for such a scenario is that of a dilaton, the (pseudo)-Goldstone boson of a spontaneously broken conformal symmetry~(CFT), with properties similar to those of the SM Higgs boson. Previous works~\cite{Carmi:2012in,Bellazzini:2012vz,Giardino:2013bma} show that the data collected by the LHC already disfavors the simple scenario of a pure dilaton, in which all the SM particles are affected similarly by the strong dynamics.

More generally, both a dilaton and an $SU(2)_W$ Higgs doublet may be present. The two fields can mix at low energies, resulting in two physical scalars, each with collider production and decay modes similar to those of a Higgs boson. It is necessary to understand the extent to which one can experimentally differentiate between the cases of a pure Higgs, a pure dilaton, and a mixture of the two.

The dilaton has experimental implications beyond Higgs phenomenology. In particular, under mild assumptions, the dilaton is expected to couple to Dark Matter (DM) particles in a well defined manner at low energies, with interaction strength proportional to the DM mass. Consequently, the dilaton field mediates the different interactions between the dark and the visible sector~\cite{Bai:2009ms}.

We therefore make the following simplifying assumptions:
\begin{enumerate}
\item[(1)] the DM interacts with the visible sector only through the dilaton field, and
\item[(2)] the SM and DM particles are fully embedded in the strongly coupled sector.
\end{enumerate}
With assumption~(1), a prediction for the DM signal rates in direct and indirect detection experiments may be obtained. Assumption~(2) implies that all SM particles couple to the dilaton through specific non-renormalizable interactions, allowing for a minimal set of parameters which span the theory space. These are:
\begin{itemize}
\item $m_\chi$, the DM mass,
\item $\alpha$, the dilaton-Higgs mixing angle,
\item $m_H$, the heavy scalar mass, and
\item $f$, the CFT breaking scale.
\end{itemize}
Assumption~(2) may be somewhat relaxed, if mixing arises between a weakly coupled  and a strongly coupled sector. In this case, different SM particles may carry distinct anomalous dimensions, leading to different couplings to the dilaton. This scenario, which is beyond the scope of this paper, is less constrained as it encompasses a much larger parameter space. (See, for instance, Ref.~\cite{Bellazzini:2012vz}.)

Under these assumptions, we explore the Higgs-dilaton mixing scenario as a portal between the visible and the dark sector (for a pre-LHC study of the dilaton portal, see~\cite{Bai:2009ms}). We study the various collider constraints on the scalar parameter space (masses and mixing), including LHC Higgs data, ElectroWeak Precision Measurements (EWPM) and the null searches for new scalars. The observed DM relic density, as well as the searches for DM, constrain the allowed model parameter space as a function of DM mass.

We analyze these constraints using two distinct scenarios for the DM relic density, that will be described in more detail below. We further define two limits in which the light scalar has SM-like properties: (i) the alignment limit and (ii) the decoupling limit, in which analytical approximations can be made. We show that existing data push the theory towards the decoupling limit, with the CFT scale above the TeV. We further present the predictions for upcoming direct searches for DM.

The plan of this paper is as follows. In section~2 we define the framework and specify the model parameter space. Section~3 is devoted to the experimental data that restrict the model. Specifically, in section~3.1 we study the different collider bounds, while observables related to the dark sector are analyzed in section 3.2. We conclude on section~4.

\section{2. Framework}
\label{sec:framework}

We begin with the low-energy description of the SM, a dilaton and a fermionic dark matter candidate. As explained in the introduction, the dark sector is assumed to be secluded from the visible sector, communicating only via the dilaton portal. We assume that the dilaton compensator field, $\phi_\sigma=f e^{\sigma/f}$, couples uniformly to the visible sector and the dark sector. This follows from the assumption that both the SM and the dark sector are embedded in the same strongly coupled sector and do not mix with elementary, weakly coupled fields~\cite{Bellazzini:2012vz}.

Under these assumption the Lagrangian takes the form
\beq \label{eq:Lint}
\mathcal{L}_\sigma=\frac{1}{2}\partial_\mu \sigma\partial^\mu\sigma+\frac{\sigma}{f}T^\mu_\mu+\ldots,
\eeq
with~\cite{Goldberger:2008zz}
\beqa
T^\mu_{\,\mu}=\sum_ig_i(\mu)(d_i-4){\cal O}_i(x)+\sum_i\beta_i(g)\frac{\partial}{\partial g_i}{\cal L_{\rm SM}}\,.
\eeqa
The SM Lagrangian, ${\cal L_{\rm SM}}$, is a sum of operators ${\cal L} = \sum_i g_i(\mu){\cal O}(x)$ at the scale $\mu$ with dimension $d_i = \dim[{\cal O}_i]$. We assume the contribution from the beta-functions above is small except for marginal operators. Moreover, we determine the energy-momentum tensor in the EW broken vacuum.

The physical spectrum contains two mass eigenstates: a light scalar, $h$, with $m_h\simeq 125$~GeV, and a heavy scalar, $H$:
\beqa
h=c_\alpha\phi_{\rm SM}+s_\alpha \sigma\,,\;\;\;\;H=-s_\alpha\phi_{\rm SM}+c_\alpha \sigma\,,
\eeqa
where $\phi_{\rm SM}$ is the SM excitation about $v=246$ GeV and $s_\alpha\equiv\sin\alpha$, $c_\alpha\equiv\cos\alpha$. The precise nature of the Higgs-dilaton mixing depends on the UV completion of the theory, whose origin is beyond the scope of this work. We note, however, that one realization of such a mixing arises from mixed kinetic term in the Lagrangian.\footnote{We thank Brando Bellazzini for this point.} In general, the mixing is dictated by a free parameter from the low energy perspective, expected to vanish as the CFT scale decouples.

The effective interaction Lagrangian, below the CFT breaking scale and above the top mass is
\beqa\label{eq:Lint}
\mathcal{L}_{\rm int}&&=\mathcal{L}_{\phi^3}-c_f^i\frac{m_f}{v}\phi_i\bar{\psi}_f \psi_f-c_\chi^i\frac{m_\chi}{v}\phi_i\bar \chi\chi \nonumber\\
&&+c^i_V\frac{2m_W^2}{v}\phi_iW^+_\mu W^{-\mu}+c^i_V\frac{m_Z^2}{v}\phi_iZ_\mu Z^\mu\nonumber\\
&&+c^i_g\frac{\alpha_s}{12\pi}\frac{\phi_i}{v}G_{\mu\nu} G^{\mu\nu}+c^i_\gamma \frac{\alpha_{em}}{\pi}\frac{\phi_i}{v}A_{\mu\nu} A^{\mu\nu}\,,
\eeqa
where $\psi_f$ are the SM fermions, $\phi_{1,2}=h,H$ and,
\beqa\label{eq:Cs}
c_{V}^h=&&c_{f}^h=c_\alpha+\rf s_\alpha\,, \;\;\;c_\chi^h=\rf s_\alpha\,, \no \\
c_{g}^h=&&+\frac{21}{2}\rf s_\alpha\,, \;\;\;\;\;\;\;\;\;\;\;\; c_{\gamma}^h=-\frac{11}{24}\rf s_\alpha\,,
\eeqa
with $r_f\equiv v/f\leq1$. The corresponding couplings of $H$ can be found by taking $\alpha \to \alpha + \pi/2$ above, {\it i.e.}, $c^H_X(\alpha) = c^h_X(\alpha+\pi/2)$. At low-energies the cubic scalar interactions take the form:
\beqa\label{eq:TLC}
{\cal L}_{\phi^3}&&=-\frac{1}{6}\lambda_{hhh}hhh-\frac{1}{6}\lambda_{HHH}HHH\no\\
&&-\frac{1}{2}\lambda_{hhH}hhH-\frac{1}{2}\lambda_{hHH}hHH\,,
\eeqa
where the trilinear couplings are specified in Appendix~A.

The different couplings presented in Eqs.~\eqref{eq:Cs} and~\eqref{eq:TLC} dictate the Higgs phenomenology, as well as the various processes which involve DM particles. These depend on the  four parameters
\beqa
r_f,\;s_\alpha,\;m_H,\;m_\chi\,.
\eeqa
To keep the perturbative expansion consistent, we assume $m_\chi/f<4\pi $ and $m_H/f<4\pi$. The SM limit is favored by the recent LHC Higgs data, see for example~\cite{Carmi:2012in,Desai:2013pga,Cao:2013cfa}. We expect this limit to be recovered in two distinct cases:
\begin{enumerate}
\item The alignment limit, $s_\alpha\ll1$, in which no mixing arises, regardless of the new physics scale~\cite{Gunion:2002zf}.
\item The decoupling limit, in which and the CFT scale is largely separated from the EW scale $r_f\ll1$. In that case one also expects $s_\alpha\ll1$.
\end{enumerate}
In the next sections we study the various experimental constraints on the Higgs-dilaton scenario, concentrating on these two limits. Out numerical results are shown in section~4.

\section{3. Experimental constraints}
\label{sec:Constraints}
\begin{figure}[t!]
\begin{center}
\includegraphics[width=0.49\textwidth]{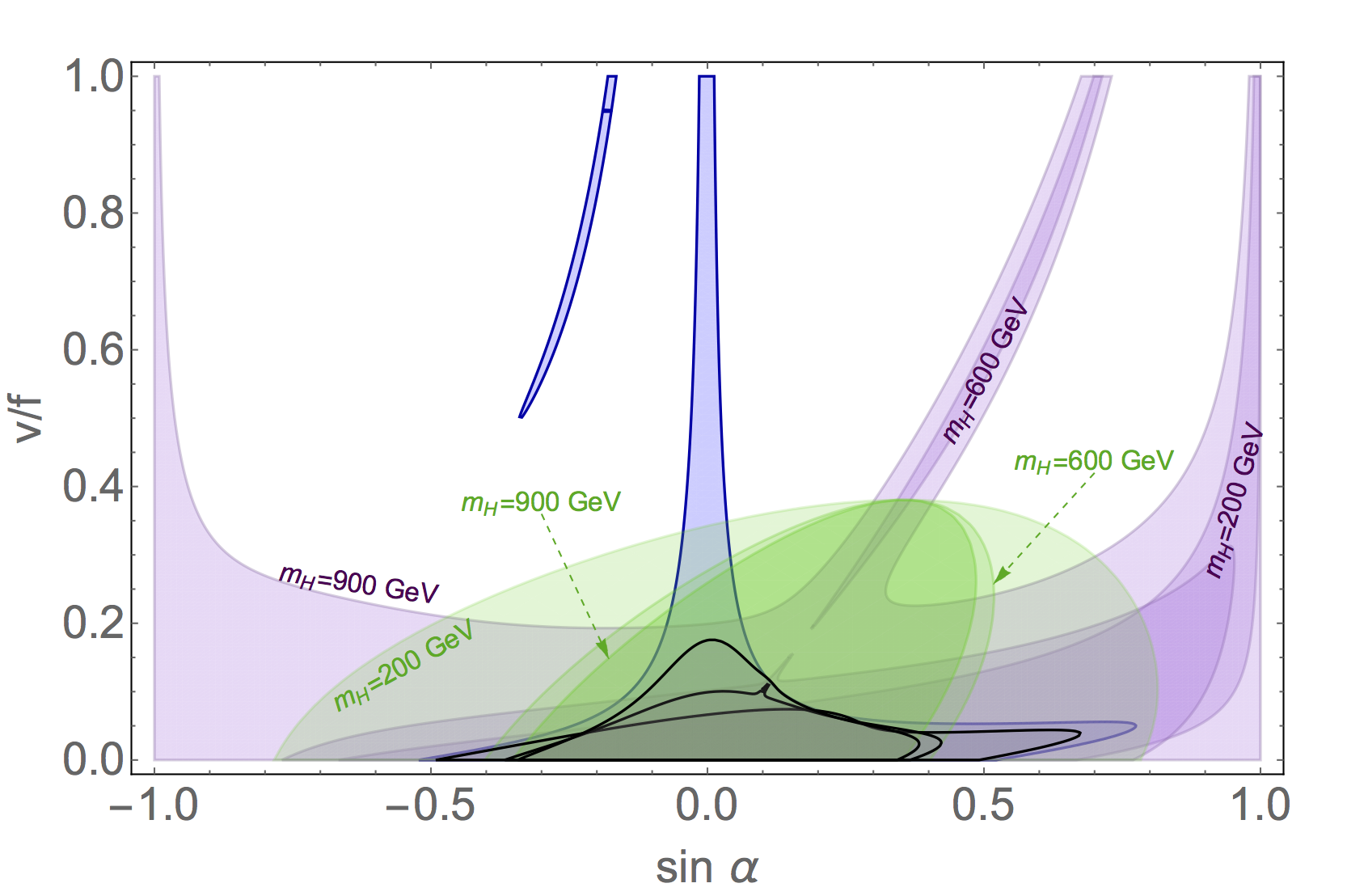}
\caption{The allowed parameter space of $r_f=v/f$ vs. $\sin\alpha$. Shown are the $2\sigma$  preferred regions by LHC Higgs searches~(blue),  EWPM~(green), and direct heavy scalar searches~(purple). The constraints from EWPM and heavy scalar searches are shown for $m_\chi = 300$ GeV and $m_H = 200,~600,~900$ GeV. Within the black boundaries are the allowed regions, combining all constraints, for each  value of $m_H$. \label{Fig:higgs}
}
\end{center}
\end{figure}

In the following we elaborate on the various experimental results which restrict the parameter space of the Higgs-dilaton model. Whenever possible, we present and discuss the analytic approximations of these constraints in the decoupling and alignment limits. Our numerical results are shown in section~4.

\subsection{3.1 Collider Constraints}\label{sec:coll}
\subsubsection{Higgs  Measurements}\label{sec:Higgs}

Much like electroweak and flavor precision measurements, the LHC Higgs rate measurements have begun to play an important role in model building.
Higgs rate measurements are reported as a confidence interval on the event rate relative to the SM prediction, denoted by $\hat{\mu}$. We consider the measured Higgs decay channels into $W^+W^-$, $ZZ$, $\gamma\gamma$, $\tau^+\tau^-$ and $b\bar{b}$ from ATLAS~\cite{dataATLAS}, CMS~\cite{dataCMS} and the Tevatron~\cite{dataTevatron}, using the SM values as taken from~\cite{Heinemeyer:2013tqa}.

Since the gluon-gluon fusion (ggF) process dominates the Higgs production at the LHC, the bulk of the Higgs production rates have similar dependence on $r_f$ and $s_\alpha$. In the decoupling and alignment limits, they read
\beq\label{eq:LHC}
\hat\mu_{ggF,h\to XX}
\simeq
\begin{cases}
1+23\rf s_\alpha-s_\alpha^2\,, & {\rm for~}X=W,Z,f\,,\\
1+22\rf s_\alpha-s_\alpha^2\,, & {\rm for~}X=\gamma\,.
\end{cases}
\eeq
An effectual estimate of the constraints can be made from a global fit combining all Higgs decay channels. One finds, at the 95\% C.L.,
\begin{equation}
\label{eq:higgsapprox}
-0.01\lesssim \rf s_\alpha\lesssim0.04\,.
\end{equation}
This result clearly shows that the LHC Higgs data push the model parameter space towards the decoupling or the alignment limits, $s_\alpha \ll 1$ and/or $\rf\ll1$. The Higgs data constraints on the Higgs-dilaton parameter space is shown in Fig.~\ref{Fig:higgs}. Note that these constraints do not depend on $m_H$ nor on the DM mass for $m_\chi\geq65$~GeV.

\subsubsection{Electroweak Precision Measurements}
\label{sec:EWPM}

In the Higgs-dilaton scenario the couplings of the light scalar to the EW gauge bosons deviate from the SM prediction at order $s_\alpha^2$. This change, along with the presence of the extra heavy scalar, modify the prediction for the oblique EW parameters with respect to their SM values~\cite{Peskin:1991sw}:
\beqa
\delta X&&=\left[\left(\rf s_\alpha+c_\alpha\right)^2-1\right]X_S(m_h)\no\\
&&+\left(\rf c_\alpha-s_\alpha\right)^2X_S(m_H)\,,
\eeqa
where $X_S$ is the scalar loop contribution to the parameter $X=S,T$, defined in Appendix~C of~\cite{Hagiwara:1994pw}. The values (and errors) for the oblique parameters obtained from the Electroweak precision measurement (EWPM) are taken from Ref.~\cite{Baak:2012kk}.

If $\rf=\tan\alpha$ is realized, the EWPM are independent of $m_H$, with $\delta X=r_f^2\, X_S(m_h)$. In this case we find that
\beqa \label{eq:rLambound}
s_\alpha\simeq r_f\lesssim0.4
\eeqa
is allowed by EWPM regardless of $m_H$. If $\rf\neq\tan\alpha$, EWPM push the parameter space of the model into both the decoupling and the alignment limits: $r_\Lambda,\;s_\alpha\ll1$. Interestingly, while the Higgs data is insensitive to $r_\Lambda$ in the alignment limit, the EW oblique parameters are still affected by the heavy scalar. The numerical results of the EWPM restrictions are shown in Fig.\ref{Fig:higgs}.

\subsubsection{Heavy Scalar Searches}
\label{sec:Hsearch}

The heavy scalar has similar production and decay channels as the light one, and therefore is tightly constrained by the LHC Higgs searches. The ATLAS~\cite{ATLASH} and CMS~\cite{Chatrchyan:2013yoa} collaborations null searches in the $W^+W^-$ and $ZZ$ decay mode place stringent bounds on the model parameter space for $m_H\leq1$~TeV. Much as with the EW precision constraints, these bounds are weakened when $\rf=\tan\alpha$, where the heavy-scalar tree-level couplings to fermions and EW bosons vanish.

A comment is in order concerning the width of the heavy scalar. While a heavy Higgs is often predicted to be rather broad, in the scenario at hand it is not necessarily so. Indeed, one finds
\beqa
\Gamma_H^{\rm tot}\simeq \left(c_{V}^H\right)^2\Gamma_{\rm SM}^{\rm tot}\left(m_H\right)+\Gamma_{H\to\bar\chi\chi}+\Gamma_{H\to hh}\,,
\eeqa
so that the heavier scalar can appear as a wide or a narrow resonance. In the numerical analysis described in section~4 we consider both options (see Ref.~\cite{ATLASH}).

The $hh$ decay channel brings another interesting possibility for a direct $H$ search. Both CMS~\cite{CMS:2014ipa} and ATLAS~\cite{Aad:2014yja} search for resonances in the $X\to hh\to b\bar{b}\gamma\gamma$ spectrum with null results. However, the resulting bounds are weak and give no additional constraints.

\subsection{3.2 Dark Matter Constraints}

\begin{figure}[t!]
\begin{center}
\includegraphics[width=.46\textwidth]{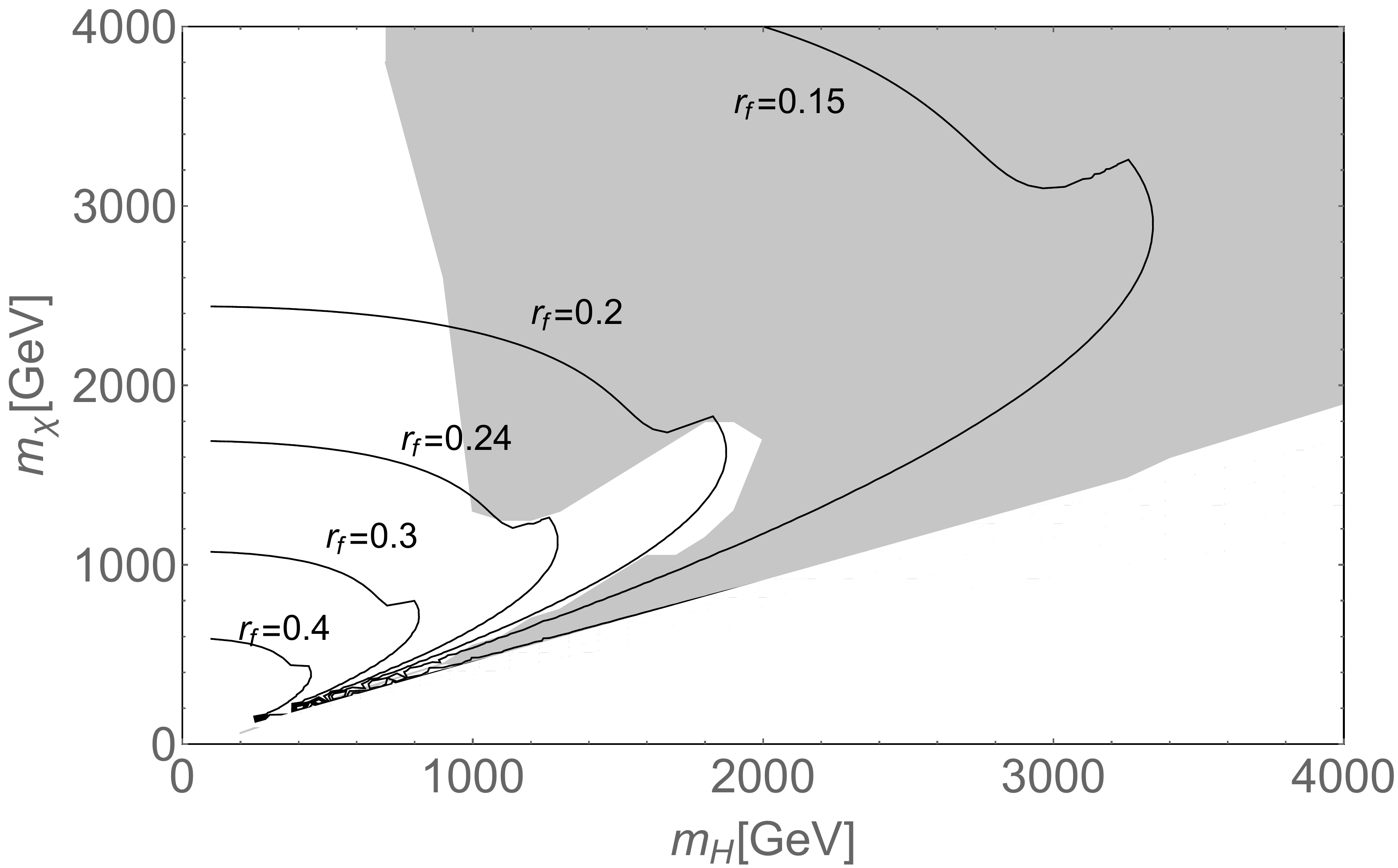}
\caption{The allowed region in the $m_H-m_\chi$ plane for $\Omega^{\rm (SM)}_{\chi}<\Omega_{\rm DM}$, including collider and direct detection constraints.  The contours show the minimal value of $r_\Lambda$ needed to avoid an overabundance of DM particles in the alignment limit. The narrow region around $m_\chi \simeq m_H/2$ is the  result of the resonant s-channel enhancement of the annihilation cross-section.
\label{Fig:darkmatterI}}
\end{center}
\end{figure}

As described above, the dilaton couples to the DM particles with interaction strength proportional to its mass, $m_\chi$. Assuming no other mediation between the dark sector and the visible
sector, this determines the expected interaction rate in direct and indirect DM detection experiments. In this section we specify the constraints arising from these searches. We further analyze the
DM annihilation processes into the visible sector which determine its relic abundance.
We consider two distinct scenarios:
\begin{itemize}
\item[(i)] {\bf $\chi$ may have hidden annihilation channels.}  $\chi$ constitutes the full DM relic abundance, while it may annihilate not only to SM particles (via the dilaton portal) but also to other dark sector states. Consequently, we assume that its relic density resulting from DM annihilations into SM particles obeys
$\Omega_\chi^{\rm (SM)}\geq\Omega_{\rm DM}$, where $\Omega_{\rm DM}h^2=0.1199\pm0.0027$~\cite{Ade:2013zuv}.
\item[(ii)] {\bf $\chi$ particles annihilate only via the dilaton portal into the SM.}  We do not demand that $\chi$ is the only DM particle, but require that it does not over close the Universe. Consequently, we take $\Omega_\chi = \Omega^{\rm (SM)}_\chi\leq\Omega_{\rm DM}$ as the only demand, and assume that the rest of the dark density arise from the relic of other, unknown particles.
\end{itemize}

As before, we first elaborate on the different constraints, and give, whenever enlightening, analytic approximations in the decoupling limit and the alignment limit. The complete numerical results are given in section~4.

\subsubsection{Relic Abundance}\label{sec:Omega}

The DM particles are in thermal equilibrium at the early Universe and annihilate into SM fermions, gauge bosons and Higgs pairs via $p$-wave processes. The total annihilation cross-section, $\sigma_{\rm ann} v= bv^2$, yields the following approximate relic abundance~\cite{KolbTurner}:
\beq\label{eq:Omegaapp}
\left(\frac{\Omega^{\rm (SM)}_{\chi}h^2}{0.12}\right)\simeq\left(\frac{5.7\times10^{-36}{\rm~cm}^2}{b}\right)\,.
\eeq
The expressions for the different annihilation modes are presented in Appendix~B. When kinematically allowed, the dark matter predominantly annihilates into $W^+W^-$ and $Z$ boson pairs. Off the $H$ resonance,
\beqa\label{eq:b_dec}
b^{\rm off}_W+b^{\rm off}_Z&\simeq&\frac{9\rf^4 m^2_\chi}{512v^4\pi (1-m_H^2/4m_\chi^2)^2}   \left(1-\frac{s_\alpha}{r_f}\frac{m_H^2}{4m_\chi^2}\right)^2
\no\\
&\simeq& 6.0\times10^{-38}\left(\frac{m_\chi}{1~{\rm TeV}}\right)^2\left(\frac{r_f}{0.1}\right)^4{\rm cm}^2\,.
\eeqa
For the two DM scenarios we consider, we find that
\begin{eqnarray}
\label{eq:LHC}
m_\chi r_f^2\lesssim100{\rm ~GeV}\,,&{\rm for~}\Omega_{\chi}^{\rm (SM)}\geq \Omega_{\rm DM}\,,\\
m_\chi r_f^2\gtrsim100{\rm ~GeV}\,,&{\rm for~}\Omega_{\chi}^{\rm (SM)}\leq\Omega_{\rm DM}\,,
\end{eqnarray}
should be held.
The above shows no limit on the first scenario in the decoupling limit. The second case requires the DM to be rather heavy. When including collider constraints, we find that
\beq\label{eq:RelicConditionII}
m_\chi\gtrsim 1.2{\rm~TeV} ~~~~~~~~\rm{off~ the~ H ~resonance},
\eeq
should be maintained in order to avoid an overabundance of dark matter.

The annihilation via the heavy scalar resonance, $m_\chi \simeq m_H/2$, plays an important role when requiring $\Omega_{\chi}^{\rm (SM)}\leq\Omega_{\rm DM}$. One finds,
\beq\begin{array}{lll}\label{eq:b_dec_on}
&&b^{\rm on}_W+b^{\rm on}_Z\sim\frac{9}{2048\pi}\frac{m^4_H}{v^4\Gamma_H^2}r_f^2\left(r_f-s_\alpha\right)^2\\
&&\simeq 1.5\times10^{-36}\left(\frac{r_f}{0.1}\right)^2\left(\frac{r_f-s_\alpha}{0.1}\right)^2\left(\frac{m_H}{1~{\rm TeV}}\right)^2\left(\frac{0.1}{\gamma}\right)^2{\rm cm}^2\,,
\end{array}\eeq
where $\gamma\equiv\Gamma_H/m_H$. However, close to the pole and when $\gamma \ll 1$, one cannot expand the cross-section in small velocity \cite{Griest:1990kh}. For all relic abundance calculations, we take the full, thermally averaged cross-section.   The resonant annihilation can allow for sufficient depletion of the DM abundance in the early Universe. When annihilating on resonance, the bound~\eqref{eq:RelicConditionII} does not hold, and we find no lower bound on the DM mass from thermal freezeout. The resonant region is clearly seen in Fig.~\ref{Fig:darkmatterI}, which shows our numerical results in the $m_\chi-m_H$ plane. The numerical calculation of the relic abundance is performed using {\tt MadDM}~\cite{Backovic:2013dpa} with {\tt MadGraph5}~\cite{Alwall:2014hca}.

A comment is in order regarding possible enhancement of the annihilation processes. Ladder diagrams with $h$ or $H$ exchange might lead to a large Sommerfeld enhancement of the different annihilation cross sections~\cite{ArkaniHamed:2008qn,Slatyer:2009vg,Agashe:2009ja}. However, in the mass region we consider here (below 4 TeV) these do not affect our conclusions. The enhancement can be significantly stronger for higher masses regime and influence both the relic abundance and indirect constraints. More details can be found in Appendix~C.

\subsubsection{Direct Detection}\label{sec:direct}
Ongoing direct detection experiments are currently probing the Higgs-mediated elastic scattering of DM off nuclei. In the scenario at hand, the elastic scattering is mediated both by the light and the heavy scalars, allowing for the existing and upcoming experiments to provide a non-trivial test of the mixed Higgs-dilaton scenario. Under our assumptions, the effective interactions~\cite{Shifman:1978zn}
\beq
\mathcal{L}^{\rm eff}_{\phi nn}=-\frac{m_n}{f}\sigma\bar nn-c_n\frac{m_n}{v}\phi_{\rm SM}\bar nn\,,
\eeq
mediate the DM-nucleon ($n$) spin-independent scattering, with $c_n\simeq0.3$~\cite{Cheung:2012qy}. The scattering cross section is given by
\beqa
\sigma_{\chi n\rightarrow \chi n}\simeq&&\frac{1}{\pi}\frac{m_n^2m_\chi^2}{v^4}\mu_n^2 r^2_f\times\no\\
&&\left[\frac{s_\alpha \left(c_n c_\alpha+r_f s_\alpha\right)}{m_h^2}-\frac{c_\alpha\left(c_n s_\alpha-r_f c_\alpha\right)}{m_H^2}\right]^2\no\\
\simeq&&1.4\times10^{-45}{\rm ~cm^2}\left(\frac{m_\chi}{1~{\rm TeV}}\right)^2\left(\frac{r_f}{0.1}\right)^2\times\no\\
&&\left[\left(\frac{s_\alpha}{0.1}\right)+0.05\left(\frac{1~{\rm TeV}}{m_H}\right)^2\left(\frac{r_f}{0.1}\right)\right]^2\,,
\eeqa
where $\mu_n=m_{\chi}m_n/(m_{\chi}+ m_n)$ is the nucleon-DM reduced mass. The last approximation is made in the decoupling limit.

\begin{figure}[t!]
\begin{center}
\includegraphics[width=.46\textwidth]{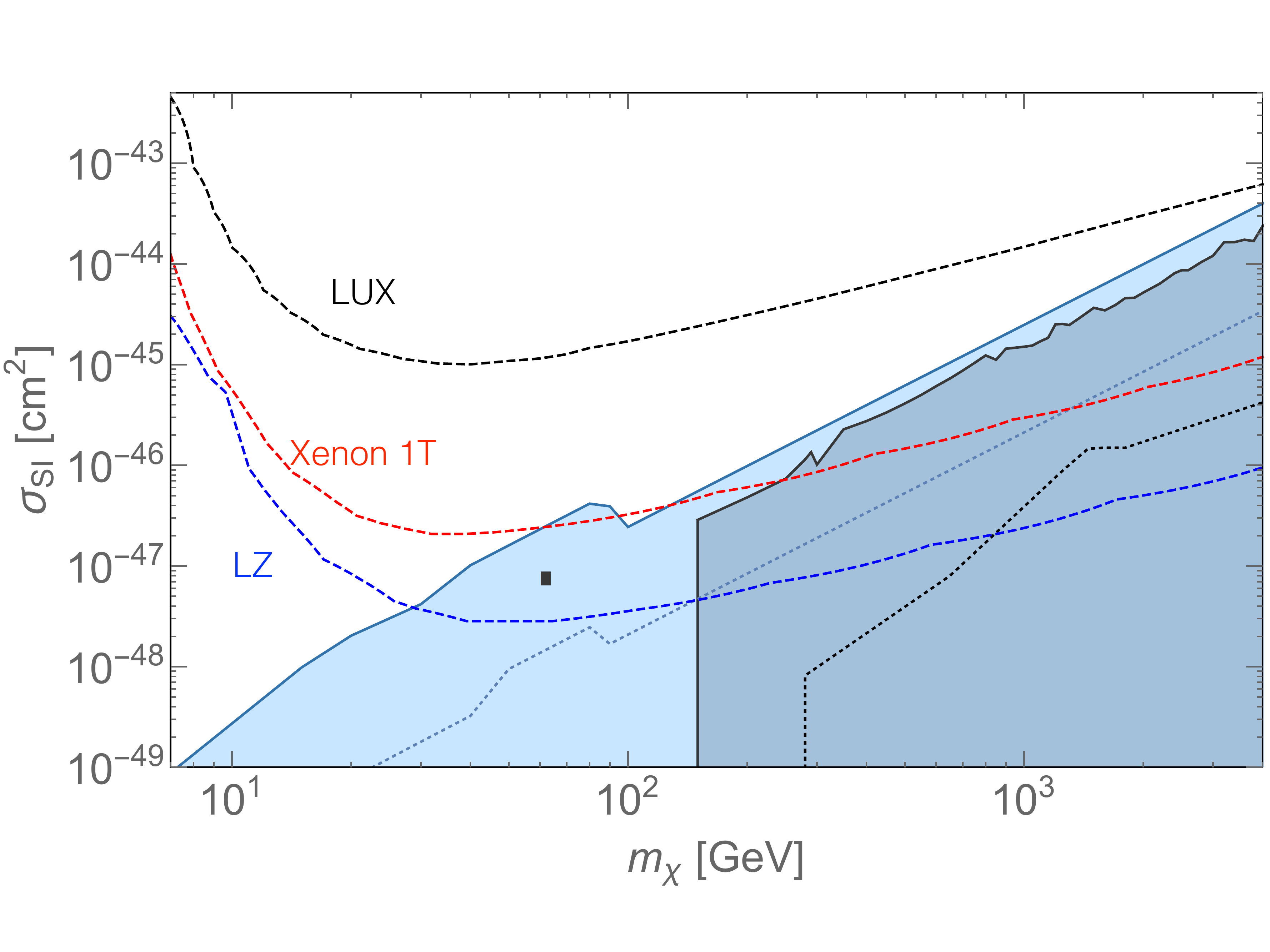}
\caption{The allowed region in $m_\chi-\sigma_{\chi n \to \chi n}$ plane for $\Omega^{\rm (SM)}_{\chi}>\Omega_{\rm DM}$ (blue) and $\Omega^{\rm (SM)}_{\chi}<\Omega_{\rm DM}$ (gray), taking into account the collider constraints.  For each scenario, we show the maximum scattering cross-section for the general case (upper line, solid) and for s=0 (lower line, dotted).   The current LUX bound (dashed black) and future Xenon 1T (dashed red) and LZ (dashed blue) bounds are also shown.
\label{Fig:darkmatterII}}
\end{center}
\end{figure}

To date, the strongest constraints arise from the null results reported by the LUX experiment~\cite{Akerib:2013tjd}. Denoting the mass-dependent experimental upper limit on the cross-section by $\sigma_n^{\rm LUX}$, one has,
\beqa
\frac{\Omega^{\rm (SM)}_\chi}{\Omega_{\rm DM}} \sigma_{\chi n\rightarrow \chi n}\leq\sigma_n^{\rm LUX}\,,
\eeqa
where $\Omega^{\rm (SM)}_\chi = \Omega_{\rm DM}$ for scenario~(i) while it may be smaller in scenario~(ii). As can be seen in Fig.~\ref{Fig:darkmatterII}, in both cases current sensitivity adds no additional constraint to the corresponding parameter space. Upcoming experiments, however, such as Xenon1T~\cite{Aprile:2012zx} and LZ~\cite{Malling:2011va} are expected to probe these models in the near future. We show these future limits, using the expected sensitivities discussed in~\cite{Cushman:2013zza}.
\subsubsection{Indirect Detection}\label{sec:indirect}

We now discuss the constraints arising from the various searches for indirect signals of DM. Local $\chi$ annihilations into the visible sector can be detected by gamma-ray telescopes, in particular, by the Fermi-LAT~\cite{Ackermann:2013yva} and the H.E.S.S~\cite{Abramowski:2011hc,Abramowski:2013ax} experiments. However, we find that for $v\simeq10^{-3}$ the total annihilation cross section is always smaller than $10^{-30}~{\rm cm}^3/{\rm sec}$ for $m_\chi\leq4$~TeV, implying no constraints from these searches. The presence of the Sommerfeld enhancement does not alter this conclusion. However, the Sommerfeld enhancement does significantly increase the annihilation rate for heavier dark matter, not shown here.

Additional constraints, derived from the CMB power spectrum~\cite{Ade:2013zuv,Hinshaw:2012aka}, may arise from the change in the ionization history due to $\chi$ annihilations in the early Universe. We find, however, that no additional bounds are imposed due to the velocity-suppressed annihilation rate and the small DM velocities at the reionization epoch. For DM lighter than 4 TeV, indirect detection experiments give weak constraints and do not affect the model parameter space allowed by other experimental bounds. These conclusions are expected to change significantly for heavier dark matter.

\section{4. Results and Discussion}
\label{sec:results}

We analyze the Higgs-dilaton scenario as a portal between dark matter and the visible sector. Various experimental results, arising from both collider studies and DM searches, bound the model parameter space and affect its predictions for future DM searches. Let us summarize out main findings. Fig.~\ref{Fig:higgs} shows the various collider constraints in the $s_\alpha-r_\Lambda$ plane, for $m_H=200,600,900$~GeV and $m_\chi=300$~GeV. These are obtained at the 95\% C.L. using a simple $\chi^2$ minimization of the LHC Higgs rates, EWPM and the direct $H$ searches. As discussed previously, while the Higgs rate observables are insensitive to the CFT breaking scale in the alignment limit, the oblique parameters might still deviate from their SM prediction. When combined together, the collider bounds are satisfied for $r_f\leq0.24$, for all values of $m_H$. The best fit point lies in the extreme decoupling limit, $s_\alpha,r_\Lambda=0$ (for $m_H=1$~TeV).

The bounds from collider studies are combined with the requirement that the $\chi$ annihilations to the SM sector are sufficient to ensure no overclosure, $\Omega_\chi^{\rm (SM)}h^2\leq0.12$. Fig.~\ref{Fig:darkmatterI} shows the viable mass range for the DM particle and the heavy scalar, in this case. We find that $m_H\geq900$~GeV and $m_\chi\geq1.2$~TeV should be realized, unless the DM annihilates via the heavy Higgs resonance. In the latter case, it is likely that the DM mass scale is related to the explicit breaking of the CFT. We note that these conclusions hold both in the alignment limit and for $s_\alpha\neq0$.

Opposite to the collider constraints, a lower bound on $r_\Lambda$ arises when considering $\Omega_\chi^{\rm (SM)}h^2\leq0.12$. This bound, drawn in the $m_\chi-m_H$ plane, can be seen in the contours of Fig.~\ref{Fig:darkmatterI}, which show the minimal values of $r_\Lambda$ needed to avoid an overabundance of DM particles. Away from the alignment limit, future Higgs precision measurements will be sensitive to much of the unconstrained parameter space. Finally, if one allows for other unknown annihilation modes for $\chi$, that deplete the relic abundance, we find no restrictions for $m_\chi$ and $m_H$.

Our predictions for the coherent DM scattering off nuclei, probed in direct detection searches, are presented in Fig.~\ref{Fig:darkmatterII}, along with the current LUX bound and future prospect of Xenon~1T and LZ sensitivities~\cite{Cushman:2013zza}. As can be seen, the Higgs-Dilaton scenario evades current direct-detection searches, but future experiments may allow for the discovery of DM.
~\\

{\bf Note added:} During the preparation of this work we became aware of Ref.~\cite{csaba} which consider a similar scenario. \\

{\bf Acknowledgments.}
We thank Mihalio Backovic, Brando Bellazzini and Gilad Perez for useful discussions. A.E., E.K., Y.S. and T.V. congratulate S.N for his 75th birthday.  T.V. and E.K. are supported in part by a grant from the Israel Science Foundation. E.K. is supported in part by the NSF grant PHY-1316222. T.V. is further supported by the US- Israel Binational Science Foundation, the EU-FP7 Marie Curie, CIG fellowship and by the I-CORE Program of the Planning Budgeting Committee and the Israel Science Foundation (grant NO 1937/12).

\appendix

\section{A. Scalar trilinear couplings}

To compute the trilinear coupling, we start from the following Lagrangian:
\beqa\label{eq:TLC}
{\cal L}_{\phi^3}&&=-\frac{1}{6}\lambda_{\rm SM}\phi_{\rm SM}^3-\frac{1}{6}\lambda_{\sigma}\sigma^3-\frac{m_{\rm SM}^2}{f}\sigma\phi_{\rm SM}^2\,,
\eeqa
where $\lambda_{\rm SM}=3m_{\rm SM}^2/v$ and $\lambda_{\sigma}=\xi m_\sigma/f$. $\xi$ is a model dependent parameter
expected to be order unity~
\cite{Goldberger:2008zz}. Rotating to the physical states
\beqa \label{eq:TLC}
\lambda_{hhh}&&=\left(\frac{M_1^2}{v}\right)3c_\alpha^2\left(c_\alpha+s_\alpha r_f\right)+\left(\frac{M_2^2}{v}\right)\xi s_\alpha^3 r_f\,,\no\\
\lambda_{HHH}&&=\left(\frac{M_1^2}{v}\right)3s_\alpha^2\left(-s_\alpha+c_\alpha r_f\right)+\left(\frac{M_2^2}{v}\right)\xi c_\alpha^3 r_f\,,\no\\
\lambda_{Hhh}&&=\left(\frac{M_1^2}{v}\right)c_\alpha\left(c_\alpha^2 r_f-3c_\alpha s_\alpha-2s_\alpha^2r_f\right) \no\\
&&+\left(\frac{M_2^2}{v}\right)\xi c_\alpha s_\alpha^2 r_f\,,\no\\
\lambda_{HHh}&&=\left(\frac{M_1^2}{v}\right)s_\alpha\left(-2c_\alpha^2 r_f+3c_\alpha s_\alpha+s_\alpha^2r_f\right) \no\\
&&+\left(\frac{M_2^2}{v}\right)\xi s_\alpha c_\alpha^2 r_f\,,
\eeqa
with
\beqa
M_1^2\equiv&&m_h^2c_\alpha^2+m_H^2s_\alpha^2\,,\;\;\;M_2^2\equiv m_h^2s_\alpha^2+m_H^2c_\alpha^2\,.
\eeqa
For our numerical results we use $\xi=5$, and verify that our final results change only little for other choices of this parameter.

\section{B. DM annihilation processes}\label{app:ann}

DM annihilation cross section are mediated by the light and heavy scalars. For all of these processes we find $a=0$, and the following $p$-wave coefficients:
\beqa\label{eq:b}
b_f=&&\frac{N_f}{8\pi}\frac{m_\chi^4m_f^2}{v^4}\beta_f^3\left|\frac{c_\chi^hc_f^h}{P_h^2}+\frac{c_\chi^Hc_f^H}{P_H^2}\right|^2\,,\\
b_V=&&\frac{g_V}{64\pi}\frac{m_\chi^2m_V^4}{v^4}\left(2+\left(1-2\frac{2m_\chi^2}{m_V^2}\right)^2\right)^2\beta_V\times\no\\
&&\left|\frac{c_\chi^hc_V^h}{P_h^2}+\frac{c_\chi^Hc_V^H}{P_H^2}\right|^2\,,\no\\
\eeqa
with,
\beqa
\beta_X&&=\sqrt{1-\frac{m_X^2}{m_\chi^2}}\,,\\
P_\phi^2&&=4m_\chi^2-m_\phi^2+i \Gamma_\phi m_\phi \,,
\eeqa
$g_W=1$ and $g_Z=1/8$. As for the scalar modes, the annihilation processes are mediated via a $t$, a $u$ and two $s$ channels. In the limit $s_\alpha=0$ these obey:
\beqa
b_h&&=\frac{r_\Lambda^4}{128\pi}\frac{m_h^4m_\chi^2}{v^4}\frac{1}{\left|P_H^2\right|^2}\beta_h\,,\no\\
b_H&&=\frac{25r_\Lambda^4}{128\pi}\frac{m_H^4m_\chi^2}{v^4}\frac{1}{\left|P_H^2\right|^2}\beta_H\,.
\eeqa

\section{C. Sommerfeld Enhancement}

The ladder scalar exchange diagrams can largely enhance the DM annihilation cross section, via the Sommerfeld enhancement~\cite{ArkaniHamed:2008qn}. For each scalar, there is an induced Yukawa potential given by $V=- \frac{\alpha_i}{r} e^{-m_{\phi_i} r}$~\cite{ArkaniHamed:2008qn,Slatyer:2009vg,Agashe:2009ja} where
\beqa
\alpha_i&=&\left(\frac{m_\chi}{v}\right)^2\frac{\left(c_\chi^{i}\right)^2}{4\pi}\,.
\eeqa
The enhancement depends on two variables
\beq
\epsilon_v \equiv \frac{v}{\alpha} ~~{\rm and}~~\epsilon_\phi \equiv \frac{m_\phi}{\alpha m_\chi }
\eeq
and is significant when $\epsilon_v,\epsilon_\phi < 1$. Schematically, this corresponds to requiring that the DM not escape the Yukawa potential well while the Yukawa range is long enough to contain the dark-matter. Since the dilaton coupling scales with $m_\chi$, the Sommerfeld enhancement becomes significant for heavy dark matter. In the parameter space we consider here, the enhancement can be as large as $10^{2}$ for the p-wave annihilation.



\end{document}